\documentstyle[stwol,aps,prl,twocolumn,psfig]{revtex}

\def\be{\begin{equation}}
\def\ee{\end{equation}}
\def\bea{\begin{eqnarray}}
\def\eea{\end{eqnarray}}
\begin{document}

\title{TOWARDS A SOLUTION
OF THE COSMOLOGICAL DOMAIN WALLS PROBLEM}

\author{ZYGMUNT LALAK}

\address{Institute of Theoretical Physics\\
University of Warsaw, 00-681 Warsaw}

\twocolumn[\maketitle\abstracts{
We show\cite{it1,lalaknew}
that all kinds of biasing of cosmological phase transitions 
produce qualitatively new type of domain wall networks.
The biased networks consist of compact, finite size, bag-like
wall structures and exhibit a generic instability. 
The surface of biased networks disappears exponentially fast 
after a limited period of scaling.
We argue that fluctuations of the background make 
the network unstable even in the case of the ``symmetric on the average''
initial distribution. We observe that the variation in parameters of the 
potential, like its hight, can influence the lifetime  of the wall 
network, contrary to the standard beliefs. 
}]

\section{Introduction}
Topological defects provide cosmologists with a set of intriguing  
mechanisms for structure formation which are quite different 
in nature to the standard inflationary paradigm.  Defects form 
at spontaneous symmetry breaking phase transitions in the early universe, 
and their subsequent field ordering dynamics can perturb the  
matter and radiation content of the universe, leaving characteristic 
signals in both the present day matter distribution and 
the microwave background. 
 
The nature of the defect is determined by the topology of the  
vacuum manifold following the phase transition.  A ${\cal{Z}}_{2}$  
manifold  with 
two disconnected vacua leads to domain walls, an ${\cal S}^1$ 
manifold to cosmic strings, an ${\cal S}^2$ manifold to 
monopoles, and an ${\cal S}^3$ manifold to cosmological texture. 
Much recent attention has focused on the gauged cosmic string, 
and global texture scenarios.  However, cosmological domain walls  
have long been considered unworkable. Zel'dovich, Kobzarev and Okun 
noticed in 1975 that the energy density of a domain wall 
network will eventually come to dominate that of matter or radiation. 
More recent attempts to save domain 
wall scenarios, such as the the so-called ``late-time'' phase  
transitions, are now known to be 
in conflict with observations of the microwave background. 
 
The most complete 
study of the dynamics of domain wall networks was given by Press, Ryden and 
Spergel\cite{press} (PRS).  
These authors studied the time evolution, during the era of matter  
domination, of topological domain wall kinks in a scalar field 
with a potential energy possessing two minima, both degenerate in energy. They 
showed conclusively that such networks rapidly evolved into long domain walls 
stretching across the universe whose surface area, and, hence, energy density, 
persisted for a long time. This persistence, or scaling 
behavior, led both to the relatively rapid domination of the energy density 
of the universe by these walls and to large distortions in the cosmic 
microwave background. Both of these results are incompatible with  
observations. 
 
It did appear, therefore, that domain walls could not have formed in the early 
universe. However, these results were based on very special
 assumptions about the initial conditions of the domain wall 
network. To be precise, these authors initialized their networks by allowing 
the computer to randomly choose, at any point on the   
lattice, either one vacuum state or the other {\it with each vacuum weighted  
with equal probability}. 
However, if, for some reason, the two vacua were to be given 
different, or biased, probabilities, then these conclusions could be 
dramatically altered.
Furthermore, it has become clear in 
recent years that non-equilibrium phase transitions, which can occur in 
realistic models of the early universe, generically lead to a biased 
choice of vacuum state. There seems to be every reason then to restudy the 
case of the walls with the modification that the 
domain wall network be initialized using biased vacuum probabilities. 
\section{Biased Transitions}
To illustrate the way the biased phase transitions can actually occur
we shall discuss in some detail the post-inflationary nonthermal 
phase transitions in a weakly coupled, out-of-equilibrium scalar 
field\cite{it2,it3}.
The inflationary epoch which we presume to precede the radiation dominated 
(RD) epoch is here the usual large-scale inflation with the Hubble parameter 
of the order of $10^{14} GeV$\footnote{It turns out that one can produce 
in a very similar way nonequilibrium transitions in fields which are coupled 
strongly enough to achieve thermal equilibrium - this requires 
inflationary stage occuring around the time of the 
transition - hence the scale 
of the Hubble parameter of the inflationary deSitter epoch to be comparable to 
the scale of symmetry breaking\cite{lalaknew}}.
Each scalar field which lives in the Universe 
udergoing very rapid, quasi-exponential expansion is subject to large quantum 
fluctuations at all wavelenghts, the overall magnitude of which is set 
by $<\phi>^2 \approx H_i^3 \delta \tau$, where $\delta \tau$ is the 
duration of the inflationary epoch and $H_i$ the Hubble parameter during inflation, in massless case, and $<\phi>^2 
\approx \frac{H_i^4}{m^2}$ in the case of a field of a mass $m$. 
More precisely, the magnitude of fluctuations (power spectrum) grows like
$\frac{1}{k^3}$, so becomes very large for large wavelength components of the 
field. The result is, that these large wavelength components correlate 
the observables measured for the field over distaces as large as the blown-up
inflationary horizon, while the small short-wavelength components 
decorrelate locally the measurements, amounting to small-scale inhomogeneities.
To see  implications this picture has for phase transitions, 
let us assume that the field has a double-well potential, the hight of the barrier between vacua being $V_o$ and the mass of the field - $m$.
Then, after inflation, in the RD epoch, when the causal FRW horizon 
grows larger than ${\ell}=\frac{1}{m}$, in each causal volume the local field starts 
rolling towards the nearest minimum of the potential.
If there would be no fluctuations, the initial conditions for rolling 
would be the same in all horizon volumes, and the field would assume 
the same vacuum state everywhere. However, if the 
characteristic scale of fluctuations, $H_i$, is at least comparable to 
the distance between the minima of the potential, there is a signiificant 
probability that in  randomly choosen volumes the fluctuations have caused
different minima to become local vacuum states. Now, whether the 
probability of 
choosing one, say $(+)$, vacuum over the other is $1/2$ or not,
it is decided by the position of the background (long-wavelengths) part 
of the field with respect to the symmetric point in the field space - 
the point eqally distant from both minima. To decide where the background
is located, one should realize that the background itself is also 
a stochastic field. In fact, let us divide the quantum 
field, including all the fluctuations born during inflation, into 
2 parts with respect to the scale ${\ell}$: the short-wavelength part
${\cal F}_{\ell}$ consisting of fourier components with $H_i^{-1} < \lambda 
< {\ell}$ and the long-wavelength part ${\cal B}_{\ell}$ with 
${\ell}< \lambda 
 < L$ where $L$ is the blown-up inflationary horizon. With such a 
division it 
is clear that ${\cal F}_{\ell}$ measures the difference or 
decorrelation of the 
field between two points at distance ${\ell}$ apart, or equvalently 
between ``mean'' (in the rms sense) values of the field in two boxes of 
size ${\ell}$ each. On the other hand, the field ${\cal B}_{\ell}$ 
measures the 
common part of the field, which looks like a constant contribution if 
one moves over distances ${\ell}$ and smaller. Of course, 
also ${\cal B}_{\ell}$
is in principle a quantum, or - at the level of quasi-classical description -
a stochastic field, which can be characterized by means of some 
probability distribution, or through moments of the distribution, like 
the mean or standard deviation. In first approximation, we can regard the 
${\cal B}_{\ell}$ field as a constant 
 classical background whose value is chosen according to a 
gaussian distribution with the width equal to the rms value of the 
underlying quantum field 
\begin{equation}
rms({\cal B}_{\ell}) = \frac{H_i}{2 \pi} \sqrt{\log ( \frac{L}{\ell})}
\end{equation}
Departures from this zeroth-order approximmation 
will be considered in the next 
section. 
It is clear now, that the probability that the classical position of the 
background would correspond to the equal probabilities of populating both 
vacua is close to zero, and that the choice of the vacuum in a 
postinflationary transition is in general biased, one of the vacua becomes 
more likely to be choosen than the other. 
Of course, in the region interpolating between different vacua 
a domain wall shall form with a large probability. The physics of so formed 
biased 
networks is summarized in the following section.
\section{Evolution of the Network}
The detailed procedure which has been used to perform numerical simulations,
and a simple analytical model which can be used to understand main 
qualitative features of the result have been described in reference\cite{it1}.
The focus here is to summarize the evolution of the surface energy density 
of the network of domain walls in the radiation dominated epoch
(the results in the matter dominated epoch are qualitatively identical).   
\noindent{\em Two Dimensional Simulations. } 
In this case we have studied\cite{it1} the evolution of $A/V$ as a function 
of elapsed conformal time, $\eta$, for a number of initial bias 
probabilities between $p=0.5$ and $0.4$.  
 Each simulation ran on a  
$1024\times 1024$ lattice until either no more 
domain walls were found in the box, or $\eta$ exceeded $512$ $(=L/2)$. 
 
For the $p=0.5$ case we recover the scaling properties reported 
in PRS. 
Fitting the scaling portion  
($10<\eta<100$) of the 
curve to the power law 
\begin{equation} 
A/V\propto\eta^{\bar{\nu}}, 
\end{equation} 
we find $\bar\nu=-0.88\pm0.04$. 
 
Moving away from the $p=0.5$ case, one sees a dramatic departure 
from self-similar scaling.  In each case there is an exponential 
cut-off in the ratio $A/V$ at some characteristic time.  For the cases of 
$p$ close to $1/2$, that is for $p$=0.457 and 0.45, we find that the curves 
are well fitted by a function of the form 
\begin{equation} 
\label{eq:fit} 
A/V \propto \eta^{\bar\nu} e^{-\eta / \bar{\eta}}. 
\end{equation} 
However, for the cases of $p$ near $1-p_{c}=0.407$, that is, $p=0.425$ and 
$0.4$, a simple exponential suffices: 
\begin{equation} 
\label{eq:fit2} 
A/V \propto e^{-\eta / \bar{\eta}}. 
\end{equation} 
Values for $\bar\nu$ and $\bar{\eta}$, 
averaged over 5 runs for each value of $p$, 
are given in Table~\ref{tb:fit1}.  
 
For $p$ close to $0.5$ the domain wall network appears to enter 
a quasi-scaling regime in which $A/V$ scales $\propto \eta^{\bar\nu}$, 
before eventually being exponentially cut-off at $\eta={\bar{\eta}}$. 
As $p\rightarrow 1-p_c=0.407$, however, $\bar{\eta}$ 
rapidly approaches the resolution size of the grid, and no 
evidence of early scaling is seen. This behavior continues as $p$ drops below 
the critical threshold, as indicated in the $p=0.4$ example. 
We find that for the cases of $p<0.4$ the exponential cut-off  
of $A/V$ becomes even more precipitous, with the characteristic scale 
$\bar{\eta}$ rapidly approaching the lattice grid size. 

To conclude, in the two-dimensional simulations we see persistent scaling 
behavior precisely at $p=0.5$. For $p$ below 0.5 but above the critical  
threshold 
$1-p_{c}=0.407$, we see scaling for a finite time which is then 
exponentially cut-off at some conformal time $\bar\eta$. The value of  
$\bar\eta$, 
which becomes very large as $p\rightarrow 0.5$, decreases rapidly as $p$ 
approaches the critical threshold. Near $p=0.407$, and below it, no scaling 
behavior is seen and the behavior is well described by a simple exponential 
for all conformal time. 
\begin{table}[p]
\begin{tabular}{|c|l|l|} 
$\qquad p\qquad$ & $\bar{\nu}\qquad\qquad$ & $\bar\eta\qquad$ 
\\ \hline
0.5 & -0.88 & -- \\ \hline
0.475 & -0.6 & 22.4 \\ \hline
0.45 & -0.68 & 7.8 \\ \hline
0.425 & -- & 3.2 \\ \hline
0.4 & --  & 2.2  \\ 
\end{tabular}
\caption{Fits to the plots of $A/V$ against $\eta$ for different
initial bias probabilities, $p$, in two dimensions,  using the
functional forms (\protect{\ref{eq:fit}}) and 
(\protect{\ref{eq:fit2}}) given in the text.}
\label{tb:fit1}
\end{table}

\noindent{\em Three Dimensional Simulations. } 
The three dimensional simulations are run on a $128^3$ grid.
Again, the self-similar evolution seen in PRS for the $p=0.5$ case 
is reproduced well in the time range $2w_0<\eta < L/2$. 
Measuring the logarithmic slope of $A/V$ versus $\eta$ between 
these times we find $\bar{\nu}=-0.89\pm 0.06$.
 
With $p\neq 0.5$, we see qualitative 
features in the three dimensional runs similar to those 
of the two dimensional cases. The one major difference is that  
the turn-over into an 
exponential decay is seen to occur much earlier, so much so 
in fact that all the cases considered here ($p=0.47,0.48,0.49$) 
are well fitted by a simple exponential curve with no initial 
pseudo-scaling regime.  Using a fitting function of the form of 
equation~\ref{eq:fit2}, the values of the $\bar\eta$ 
for each $p$ are given in Table~\ref{tb:fit2}. 
\begin{table}[p]
\begin{tabular}{|c|l|} 
$\qquad p\qquad$ &$\bar\eta\qquad$ 
\\ \hline
0.49 & 5.4  \\ \hline
0.48 & 2.8 \\ \hline
0.47 & 1.8 \\
\end{tabular}
\caption{Fits to the plots of $A/V$ against $\eta$ for different
initial bias probabilities, $p$, in three dimensions  using the
functional form (\protect{\ref{eq:fit2}}) given in the text.}
\label{tb:fit2}
\end{table}

\begin{figure*}
\begin{center}
\mbox{
\psfig{figure=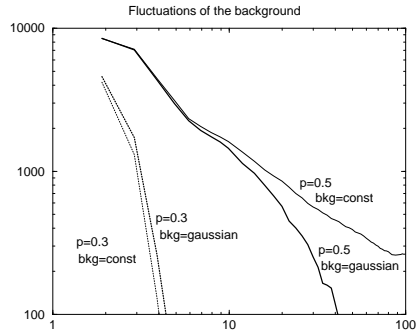,height=2.0in}
      }
\end{center}
\caption{The evolution of the surface area of the network 
when background fluctuations are present. The thin solid line corresponds 
to the constant value of $p=0.5$ while the thick solid line shows 
the situation where p fluctuates around $p_o=0.5$. Instability in the later 
case is clearly visible. The thin dotted line presents the evolution 
of the network created with constant value of $p=0.3$. The thick dotted line 
gives evolution of the biased network when $p$ fluctuates around $p_o=0.3$.
All curves correspond to d=2 dimensions. The verical axis gives surface
of the network, the horizontal axis gives conformal time. 
\label{fig:1}}
\end{figure*}

To conclude, in the three dimensional case we see persistent scaling behavior 
precisely at $p=0.5$. For any value of $p$ below 0.5, little or no scaling 
behavior is seen and the behavior is well described by a simple exponential 
for all conformal time. 
\noindent{\em Background Fluctuations\cite{lalaknew}. }
The new result which we want to discuss  here is the influence of the 
fluctuations of the background on the behaviour of the network. 
In the simulations described so far the background value of the stochastic 
field was assumed to be a constant randomly located in the field space 
(the freedom of choosing this background value is reflected in the 
freedom of choosing the bias parameter $p$). However, we know from the previous section that in fact the background is not strictly 
homogeneous, as there must be fluctuations around the zeroth-order 
approximmation constant background value.
In the simulations whose results are depicted in Figure~\ref{fig:1}
we are simulating the fluctuations of the background through 
fluctuations of the probability $p$. At each lattice site we draw 
$p$ according to the gaussian distribution with some central value
$p=p_o$ and the width equal to $\sigma = 0.25 p_o$. 
As one can see from the plot, this  has a dramatic impact on the bahaviour of 
the equilibrium network with previously constant value $p=p_o=0.5$. 
Now, with fluctuating $p$, the network becomes unstable, and 
its surface again decays quasi exponentially\cite{lalaknew}. 
Intuitively one can understand this result noticing, 
that the average value of the probability over a randomly choosen 
small subvolume of the lattice is typically different from $p_o=0.5$,
hence each small part of the network behaves like a biased network,
and as these subnetworks are unstable, the averaging over many subnetworks 
cannot produce the equilibrium bahaviour, despite the fact that 
the sub-probabilities average to $p_o=0.5$. 

\begin{figure*}
\begin{center}
\mbox{
\psfig{figure=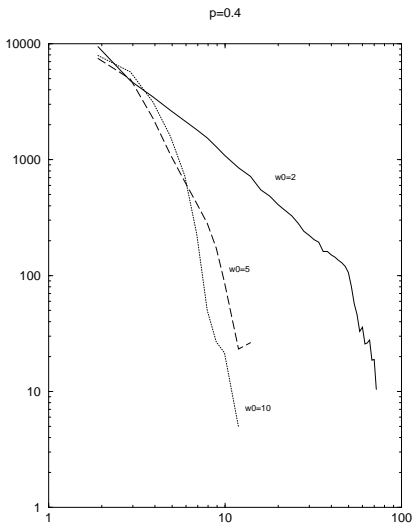,height=2.0in}
      }
\end{center}
\caption{The curves show the dependence of the evolution of the network on
the parameters of the potential through the domain wall thickness $w_o$. 
There is a tendency for the network to persist for a longer period of time 
for thinner walls. The figure shows the case of $p=4$ in d=2 dimensions. 
The results in d=3 are qualitatively the same. The verical axis gives surface
of the network, the horizontal axis gives conformal time.
\label{fig:2}}
\end{figure*}

The other case shown in the picture (dotted curves, $p=0.3$) illustrates
the general tendency of biased networks to persist for a longer period 
of time (slower decay rate) when the probability (in fact - the background
field configuration) 
fluctuates between different lattice sites.   
The fluctuating $p=0.5$ case shows that in reality all the networks 
formed during actual cosmic transitions will be biased, hence 
after a finite scaling period, typically rather short, will disappear 
dissipating their surface energy in the form of scalar (maybe also 
gravitational) waves.
\noindent{\em Wall Thickness\cite{it1,lalaknew}. }
The next observation which can have some significance for the problem of 
the possible impact of biased networks on structure formation is that
the values of $\bar\nu$ and $\bar{\eta}$ 
were found to be weakly dependent 
on the value of $w_0$, the domain wall thickness.  
Changing the wall thickness affects the 
network evolution through two closely balanced effects. 
Increasing the height of the potential barrier between the two vacua,
$V_o = \frac{1}{2} \frac{\pi^2}{w_o^2}$, 
makes transition from one vacuum to the other energetically 
less favorable.  But, this is compensated somewhat, not necessarily 
exactly, by the increased 
domain wall surface tension, $\sigma = \frac{1}{2} \frac{\pi^2}{w_o}$. 
However, for a quartic potential, 
the value of $V_0$ clearly does not scale out of the equations 
of motion, so  this balance need not be exact. 
(In the domain wall scenario, the scalar field is 
off the vacuum manifold at many points throughout space, 
making the evolution more dependent on the relative values 
of $V_0$ and $\phi_0$ than, say, a texture scenario, where  
the non-linear sigma model works well, and evolution is  
effectively independent of the potential barrier between vacua.) 
This is illustrated in the Figure~\ref{fig:2}. 
\section{Conclusions}
From the presented results it becomes clear that biased transitions 
solve the domain walls problem very efficiently and naturally 
due to the generic instability of the network with respect to the departure 
from $p=0.5$ and with respect to fluctuations of the background. 

The interesting and important for structure formation problem which remains 
to be clarified is whether there is a possibility of having unstable networks 
with scaling periods long enough to show up on the $\frac{\delta \rho}{\rho}$
curve. This would be an interesting possibility in the 
context of hot dark matter scenarios. The other point which has to be 
considered more detaily is interaction of biased networks with matter. 

Recenly we became aware of the work of S. Larsson, S. Sarkar
and P. White\cite{lsw} which confirms and extends our
invstigations of biased networks.
\section*{Acknowledgments}
This work has been supported in part by Polish Committee
for Scientific Research -- KBN Grant, and by the EEC under the 
``Flavourdynamics'' Network.

\end{document}